# Diamond electro-optomechanical resonators integrated in nanophotonic circuits

P. Rath[1,+], S. Ummethala[1,+], S. Diewald[2], G. Lewes-Malandrakis[3], D. Brink[3], N. Heidrich[3], C. Nebel[3], and W.H.P. Pernice[1,*]

[1]*Institute of Nanotechnology, Karlsruhe Institute of Technology, 76344 Eggenstein-Leopoldshafen, Germany*

[2]*Center for Functional Nanostructures, Karlsruhe Institute of Technology, 76131 Karlsruhe, Germany*

[3]*Fraunhofer Institute for Applied Solid State Physics, Tullastr. 72, 79108 Freiburg, Germany*

+ *The authors contributed equally to this work.*

(October 18th, 2014)

Diamond integrated photonic devices are promising candidates for emerging applications in nanophotonics and quantum optics. Here we demonstrate active modulation of diamond nanophotonic circuits by exploiting mechanical degrees of freedom in free-standing diamond electro-optomechanical resonators. We obtain high quality factors up to 9600, allowing us to read out the driven nanomechanical response with integrated optical interferometers with high sensitivity. We are able to excite higher order mechanical modes up to 115 MHz and observe the nanomechanical response also under ambient conditions.

[*]Email: wolfram.pernice@kit.edu

P. Rath, S. Ummethala, S. Diewald, G. Lewes-Malandrakis, D. Brink, N. Heidrich, C. Nebel, and W.H.P. Pernice

Integrated photonic devices are receiving continued interest for technological and scientific applications. While traditionally materials originating from the electronics industry were explored, nowadays alternative options with enhanced properties are pursued for advanced applications[1,2]. In particular, materials with wide electronic band gaps are investigated to enable broadband optical operations of integrated photonic devices. These include silicon nitride and III-Nitride semiconductors such as gallium nitride and aluminum nitride, which offers one of the largest band gaps of all semiconducting materials[3]. Porting optical technology from a silicon platform to alternative materials has thus enhanced the range of applications that optical devices cover to date[4]. Recently, the possibility to integrate also nanomechanical structures into photonic circuits has provided these predominantly passive structures with further degrees of freedom for tunability. This can be achieved by exploiting gradient optical forces for all-optically tunable devices[5]. However, often stronger driving forces are required which can be conveniently provided by electrostatic actuation[6,7]. To do so efficiently, not only a convenient design, but also good optical and mechanical material properties are required. In this respect diamond is an attractive material choice. In recent years there has been a growing interest in integrated optical devices out of diamond[8–11]. There has also been a surge for fabricating high-Q free-standing mechanical resonators out of diamond which is also suitable for integrated optics in order to create optomechanical resonators[12,13,14]. Diamond has also gained interest as an ideal material for quantum optics and integrated optical circuits, due to its high refractive index, its large band gap of 5.45 eV, making it transparent from far infrared to UV wavelengths and the ability to host color centers, such as the NV[15] and SiV[16] center, which can be utilized as single photon sources. With its large Young's modulus up to 1100 GPa, for single-crystalline and as well as for

P. Rath, S. Ummethala, S. Diewald, G. Lewes-Malandrakis, D. Brink, N. Heidrich, C. Nebel, and W.H.P. Pernice

polycrystalline diamond[17], mechanical resonators with high frequencies and high quality factors are achievable[18,19]. This enables optomechanical devices for a broad range of applications such as optical actuators, vibration sensors[20] and reconfigurable optical elements[21]. Adding electrodes to the chip provides further electrical degrees of freedom for actuation of the mechanical components[22,23]. In this way, coherent driving at high frequencies and high amplitudes becomes possible, while maintaining the high read-out sensitivity of on-chip optical interferometers. Furthermore, stronger driving forces can be achieved compared to optical forces in a similar geometry[24] and cross-talk can be avoided. Such electrically driven diamond optomechanical resonators could find applications as phase-shifters[23,25] for integrated optical circuits with applications in quantum information processing[26–29]. Other applications include integrated optomechanical circuits which couple mechanical motion to the spin states of color centers[30–35].

Here, we show the first implementation of diamond integrated electro-optomechanical circuits featuring sensitive interferometric motion readout and high quality factor mechanical resonators. Including electrodes enable us to excite higher order mechanical resonances up to 115 MHz both in vacuum and under ambient conditions. Our integrated electro-optomechanical circuits as shown in Figure 1a consist of three different functional parts: optical components, mechanical resonators and metal electrodes. The optical components (shown in blue) comprise of focusing grating couplers for in- and out-coupling of light, waveguides and 50/50 splitters which form a Mach-Zehnder interferometer (MZI) with a path difference of 100 µm. The free-standing mechanical resonators (shown in green) are connected to the bulk diamond layer at four clamping points, forming an "H-resonator"[23]. One arm of the mechanical resonator is evanescently coupled to one of the waveguides of the MZI, such that a change in distance between resonator and waveguide changes the propagation constant of the optical mode which is

P. Rath, S. Ummethala, S. Diewald, G. Lewes-Malandrakis, D. Brink, N. Heidrich, C. Nebel, and W.H.P. Pernice

guided inside the waveguide. This motion thus introduces an additional phase shift. The phase shift in turn changes the interference condition and leads to a change in transmitted intensity which can be detected with a fast photodetector. Two electrodes on top of the diamond layer (shown in golden color) are designed as a parallel plate capacitor, enabling us to electrostatically drive the nanomechanical motion of the H-resonator. One of the two electrodes is fabricated directly on top of the free-standing mechanical resonator, while the second electrode is located on the diamond layer which is fixed to the underlying substrate. The electrodes are separated by a small gap, ranging from 150 nm to 300 nm for different devices on the same chip. The two separate electrodes are connected to large contact pads, which allow electrical access to the devices through micro probes. The central part of the mechanical resonator contains a two-dimensional photonic crystal lattice which acts as a mirror block for light inside the free-standing photonic part of the circuit. This way guided light does not propagate into the central area of the mechanical resonator and the electrical components. Essentially photonic and electronic components are decoupled by the photonic crystal such that the metal electrodes do not lead to absorption of light.

We fabricate the devices out of polycrystalline CVD diamond thin films, deposited in a microwave reactor onto oxidized silicon wafer, resulting in a diamond on insulator (DOI) wafer as described in our earlier work[14,24]. After deposition, the diamond thin films are polished via a slurry based chemo-mechanical polishing process[36] to an RMS roughness below 3 nm on a 25 µm$^2$ area, as confirmed by atomic force microscopy. This reduced roughness allows for high-resolution electron beam lithography with good reproducibility. Our devices are prepared from a diamond on insulator (DOI) stack with 600 nm of diamond and 2 µm buried oxide on top of a silicon carrier wafer. The devices are fabricated in three steps of electron beam lithography and

P. Rath, S. Ummethala, S. Diewald, G. Lewes-Malandrakis, D. Brink, N. Heidrich, C. Nebel, and W.H.P. Pernice

several wet and dry etching steps. The first lithography step defines electrodes and markers with Cr/Au/Cr layer of thicknesses 5/100/10 nm, respectively, patterned through a PMMA lift-off. In the second e-beam exposure, the optical components and mechanical resonators are exposed using a negative resist (FOX15). This pattern is transferred into the diamond layer using capacitively coupled reactive ion etching (RIE) with O2/Ar chemistry down to 50% etch depth. The final third lithography step defines rectangular mask openings in PMMA, which are then transferred by wet etching into a Cr hard mask. A second RIE step is used to fully etch the diamond layer in the mask region and afterwards buffered oxide etchant is used for underetching the mechanical resonators, resulting in free-standing electro-optomechanical resonators. In this fashion many electro-optomechanical circuits are patterned onto the same chip.

In order to characterize the performance of the circuits, the fabricated chip is placed into a vacuum measurement setup, pressure controlled down to $10^{-6}$ mbar. The chip is mounted onto a three-axis stage with piezo actuators, which allows for aligning the device under test to optical fibers combined in a fiber array. A tunable laser (Santec TSL-510) is connected via an erbium doped fiber amplifier (EDFA) and a polarization controller to the fiber array. The wavelength of the laser is chosen such that the MZI is operated at the quadrature point for highest sensitivity. The measurements are performed with about 1 mW laser power inside each interferometer arm. After propagation through the interferometer, the output grating coupler couples the transmitted light into a second fiber which is connected to a fast low noise photodetector (New Focus 1811, 125-MHz bandwidth). An electrical micro probe is mounted onto a one axis piezo z-stage, which allows contacting electrodes on individual devices on the chip while maintaining high vacuum conditions. The output from the micro probe is connected to a bias tee, which is biased with a tunable DC voltage source. The RF port is connected to a vector network analyzer (VNA) which

P. Rath, S. Ummethala, S. Diewald, G. Lewes-Malandrakis, D. Brink, N. Heidrich, C. Nebel, and W.H.P. Pernice

also acts as the source for radio frequency (RF) signals. The VNA is operated in transmission measurement mode, with the output port connected to the bias tee and the input port connected to the fast photodetector which records oscillating light intensities due to the mechanical motion.

By applying a constant DC voltage to the on-chip electrodes a constant attractive force can be applied between the electrodes, thus pulling the mechanical resonator closer to the fixed electrode. An additionally applied RF signal gives rise to an oscillating motion of the resonator at the frequency of the applied RF signal. The leading term of the dynamic force is proportional to the product of the RF and DC voltage[37], as the RF voltage is typically small compared to the DC voltage. In order to characterize the mechanical resonator the RF frequency is swept while recording the power spectral density. At each mechanical eigenfrequency, the amplitude of the motion is resonantly enhanced giving rise to a Lorentzian peak in the spectrum. Figure 2a shows a recorded spectrum for a resonator with a length of 40 µm and an arm width of 600 nm, when 5 V DC voltage and 20 dBm RF power were applied to the electrodes. When sweeping the RF frequency from 5 MHz to 125 MHz more than 20 resonances can be easily detected, corresponding to the different eigenfrequencies of the oscillator. From a Lorentzian fit to the observed resonances we extract the quality factors for all observed resonances, as shown in Figure 2b. While for the fundamental in-plane and out-of-plane modes around 10 MHz mechanical quality factors up to 8700 are observed, the quality factor decreases with increasing frequency, while even at 115 MHz mechanical resonances with Q factors up to 1300 are measured. The highest measured quality factor of 9600 at 13.8 MHz was found for a device with a length of 40 µm and arm width of 800 nm. Resonators with the same device geometry out of the same material, but without electrodes were shown to have Q factors up to 29.000[14]. We attribute the smaller Q factors in this work to the additional damping introduced by the gold

P. Rath, S. Ummethala, S. Diewald, G. Lewes-Malandrakis, D. Brink, N. Heidrich, C. Nebel, and W.H.P. Pernice

electrodes which are sitting directly on top of the resonators. The highest measured mechanical resonance frequency in this work is 115 MHz, measurement limited by the bandwidth of the detector of 125 MHz. By scaling down the length of the resonators and using a faster photodetector higher order modes could potentially be detected. Besides the Lorentzian resonance in the amplitude signal, also the phase between exiting force and mechanical motion is measured and the expected 180° phase shift around the resonance is observed, as shown exemplarily in Figure 3a for one of the resonances.

When increasing the RF power, distortions from the Lorentzian shape become clear, as shown in Figure 3b. This is due to stiffening Duffing nonlinearity as a result of the tensile stress present in the diamond layer, consistent with previous findings[24]. When reducing the RF power the motion of the resonator can be observed for driving powers as low as -40 dBm.

For sensing applications it is advantageous to operate the electro-optomechanical resonators under non-vacuum conditions. We therefore characterize the mechanical resonance at various pressures inside the vacuum chamber and extract the mechanical quality factor from Lorentzian fits to the power spectral density, as shown in inset of Figure 4. For the studied device the mechanical quality factor for the eigenmode at 10.8 MHz is about 8400 at low pressures down to $2.4*10^{-6}$ mbar. When increasing the pressure over several orders of magnitude from $2.4*10^{-6}$ mbar to 1.5 mbar the observed resonance shape and the extracted quality factor do not change. For pressures above 1.5 mbar up to ambient pressure the air damping is dominating the other damping mechanisms and the quality factor steadily decreases, comparable to previous studies of mechanical resonators at varying pressures[38,39]. At ambient pressure the driven mechanical motion can still be observed and the quality factor accounts to 210.

P. Rath, S. Ummethala, S. Diewald, G. Lewes-Malandrakis, D. Brink, N. Heidrich, C. Nebel, and W.H.P. Pernice

In conclusion we have demonstrated the first integrated electro-optomechanical circuits made from diamond. Utilizing electron beam lithography to structure wafer scale deposited CVD diamond thin films nanomechanical resonators with quality factors up to 9600 at MHz frequencies are achieved. Integrating the mechanical components into diamond photonic circuits allows for sensitive motion readout via on-chip interferometers. Metal electrodes patterned right on the resonators allow to drive higher order mechanical modes up to 115 MHz, while actuation and motion detection is shown at ambient conditions. By adjusting the geometry together with the actuation via electrodes such devices could potentially be used for driven motion at even higher frequencies, where air damping would not play a significant role[40].

W.H.P. Pernice acknowledges support by the DFG grants PE 1832/1-1 & PE 1832/1-2 and the Helmholtz society through grant HIRG-0005. P. Rath acknowledges support by the Deutsche Telekom Stiftung. S. Ummethala and P. Rath acknowledge support by the Karlsruhe School of Optics and Photonics (KSOP). We also acknowledge support by the Deutsche Forschungsgemeinschaft (DFG) and the State of Baden-Württemberg through the DFG-Center for Functional Nanostructures (CFN) within subproject A6.4.

P. Rath, S. Ummethala, S. Diewald, G. Lewes-Malandrakis, D. Brink, N. Heidrich, C. Nebel, and W.H.P. Pernice

References:

[1] B. Jalali and S. Fathpour, J. Light. Technol. **24**, 4600 (2006).

[2] M. Smit, J. van der Tol, and M. Hill, Laser Photon. Rev. **6**, 1 (2012).

[3] C. Xiong, W.H.P. Pernice, X. Sun, C. Schuck, K.Y. Fong, and H.X. Tang, New J. Phys. **14**, 095014 (2012).

[4] A. Politi, M.J. Cryan, J.G. Rarity, S. Yu, and J.L. O'Brien, Science (80-. ). **320**, 646 (2008).

[5] K.Y. Fong, W.H.P. Pernice, M. Li, and H.X. Tang, Opt. Express **19**, 15098 (2011).

[6] S. Sridaran and S.A. Bhave, Opt. Express **19**, 9020 (2011).

[7] H. Miao, K. Srinivasan, and V. Aksyuk, New J. Phys. **14**, 075015 (2012).

[8] A. Faraon, P. Barclay, C. Santori, K.-M.C. Fu, and R.G. Beausoleil, Nat. Photonics **5**, 301 (2011).

[9] X. Checoury, D. Neel, P. Boucaud, C. Gesset, H. Girard, S. Saada, and P. Bergonzo, Appl. Phys. Lett. **101**, 171115 (2012).

[10] P. Rath, N. Gruhler, S. Khasminskaya, C. Nebel, C. Wild, and W.H.P. Pernice, Opt. Express **21**, 11031 (2013).

[11] B.J.M. Hausmann, I. Bulu, V. Venkataraman, P. Deotare, and M. Lončar, Nat. Photonics **8**, 369 (2014).

[12] L. Kipfstuhl, F. Guldner, J. Riedrich-Möller, and C. Becher, Opt. Express **22**, 12410 (2014).

[13] M.J. Burek, N.P. de Leon, B.J. Shields, B.J.M. Hausmann, Y. Chu, Q. Quan, A.S. Zibrov, H. Park, M.D. Lukin, and M. Lončar, Nano Lett. **12**, 6084 (2012).

[14] S. Ummethala, P. Rath, G. Lewes-Malandrakis, D. Brink, C. Nebel, and W.H.P. Pernice, Diam. Relat. Mater. **44**, 49 (2014).

[15] M.W. Doherty, N.B. Manson, P. Delaney, F. Jelezko, J. Wrachtrup, and L.C.L. Hollenberg, Phys. Rep. **528**, 1 (2013).

[16] L.J. Rogers, K.D. Jahnke, T. Teraji, L. Marseglia, C. Müller, B. Naydenov, H. Schauffert, C. Kranz, J. Isoya, L.P. McGuinness, and F. Jelezko, Nat. Commun. **5**, 4739 (2014).

[17] O.A. Williams, A. Kriele, J. Hees, M. Wolfer, W. Mueller-Sebert, and C.E. Nebel, Chem. Phys. Lett. **495**, 84 (2010).

[18] Y. Tao, J.M. Boss, B.A. Moores, and C.L. Degen, Nat. Commun. **5**, 3638 (2014).

[19] P. Ovartchaiyapong, L.M.A. Pascal, B.A. Myers, P. Lauria, and A.C. Bleszynski Jayich, Appl. Phys. Lett. **101**, 163505 (2012).

[20] E. Ollier, P. Philippe, C. Chabrol, and P. Mottier, J. Light. Technol. **17**, 26 (1999).

[21] P.B. Deotare, I. Bulu, I.W. Frank, Q. Quan, Y. Zhang, R. Ilic, and M. Loncar, Nat. Commun. **3**, 846 (2012).

[22] R. Perahia, J.D. Cohen, S. Meenehan, T.P.M. Alegre, and O. Painter, Appl. Phys. Lett. **97**, 191112 (2010).

[23] M. Poot and H.X. Tang, Appl. Phys. Lett. **104**, 061101 (2014).

[24] P. Rath, S. Khasminskaya, C. Nebel, C. Wild, and W.H.P. Pernice, Nat. Commun. **4**, 1690 (2013).

[25] X. Guo, C.-L. Zou, X.-F. Ren, F.-W. Sun, and G.-C. Guo, Appl. Phys. Lett. **101**, 071114 (2012).

[26] I. Bayn, B. Meyler, A. Lahav, J. Salzman, R. Kalish, B.A. Fairchild, S. Prawer, M. Barth, O. Benson, T. Wolf, P. Siyushev, F. Jelezko, and J. Wrachtrup, Diam. Relat. Mater. **20**, 937 (2011).

[27] I. Aharonovich, A.D. Greentree, and S. Prawer, Nat. Photonics **5**, 397 (2011).

[28] B.J.M. Hausmann, B. Shields, Q. Quan, P. Maletinsky, M. McCutcheon, J.T. Choy, T.M. Babinec, A. Kubanek, A. Yacoby, M.D. Lukin, and M. Loncar, Nano Lett. **12**, 1578 (2012).

[29] F. Dolde, I. Jakobi, B. Naydenov, N. Zhao, S. Pezzagna, C. Trautmann, J. Meijer, P. Neumann, F. Jelezko, and J. Wrachtrup, Nat. Phys. **9**, 139 (2013).

[30] S. Kolkowitz, A.C.B. Jayich, Q.P. Unterreithmeier, S.D. Bennett, P. Rabl, J.G.E. Harris, and M.D. Lukin, Science (80-. ). **335**, 1603 (2012).

[31] P. Maletinsky, S. Hong, M.S. Grinolds, B. Hausmann, M.D. Lukin, R.L. Walsworth, M. Loncar, and A. Yacoby, Nat. Nanotechnol. **7**, 320 (2012).

[32] S. Hong, M.S. Grinolds, P. Maletinsky, R.L. Walsworth, M.D. Lukin, and A. Yacoby, Nano Lett. **12**, 3920 (2012).

[33] E.R. MacQuarrie, T.A. Gosavi, N.R. Jungwirth, S.A. Bhave, and G.D. Fuchs, Phys. Rev. Lett. **111**, 227602 (2013).

[34] J. Teissier, A. Barfuss, P. Appel, E. Neu, and P. Maletinsky, arXiv Prepr. **1403.3405**, (2014).

[35] P. Ovartchaiyapong, K.W. Lee, B.A. Myers, and A.C.B. Jayich, arXiv Prepr. **1403.4173**, (2014).

[36] E.L.H. Thomas, G.W. Nelson, S. Mandal, J.S. Foord, and O.A. Williams, Carbon N. Y. **68**, 473 (2014).

[37] V. Kaajakari, A.T. Alastalo, and T. Mattila, IEEE Trans. Ultrason. Ferroelectr. Freq. Control **53**, 2484 (2006).

[38] D.M. Karabacak, V. Yakhot, and K.L. Ekinci, Phys. Rev. Lett. **98**, 254505 (2007).

[39] M. Li, H.X. Tang, and M.L. Roukes, Nat. Nanotechnol. **2**, 114 (2007).

[40] X. Sun, K.Y. Fong, C. Xiong, W.H.P. Pernice, and H.X. Tang, Opt. Express **19**, 22316 (2011).

P. Rath, S. Ummethala, S. Diewald, G. Lewes-Malandrakis, D. Brink, N. Heidrich, C. Nebel, and W.H.P. Pernice

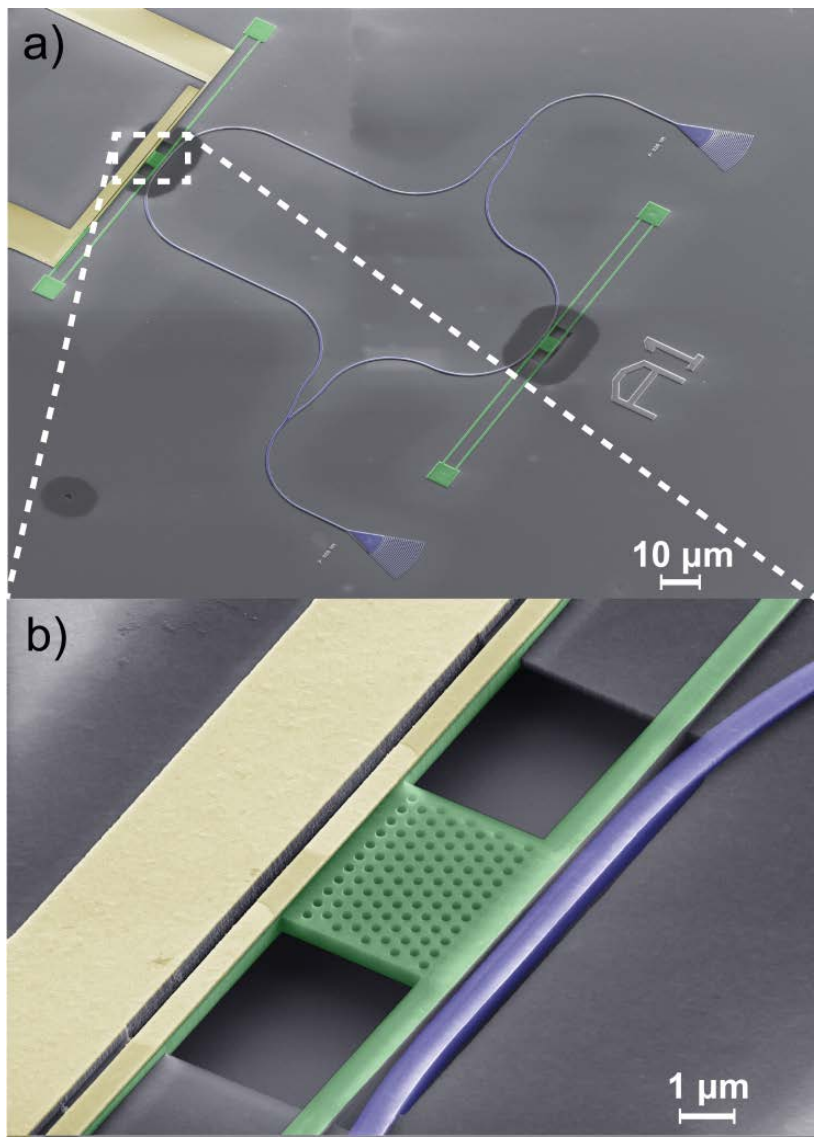

FIG. 1. False colour SEM-micrographs of a fabricated electro-optomechanical device. (a) The integrated Mach-Zehnder interferometer is shown in blue. The mechanical resonator, which is evanescently coupled to the waveguide, is shown in green, while the metal electrodes are shown in golden colour. (b) Detailed view of the free-standing resonator. The photonic crystal mirror separates the optical components from the electrode section.

P. Rath, S. Ummethala, S. Diewald, G. Lewes-Malandrakis, D. Brink, N. Heidrich, C. Nebel, and W.H.P. Pernice

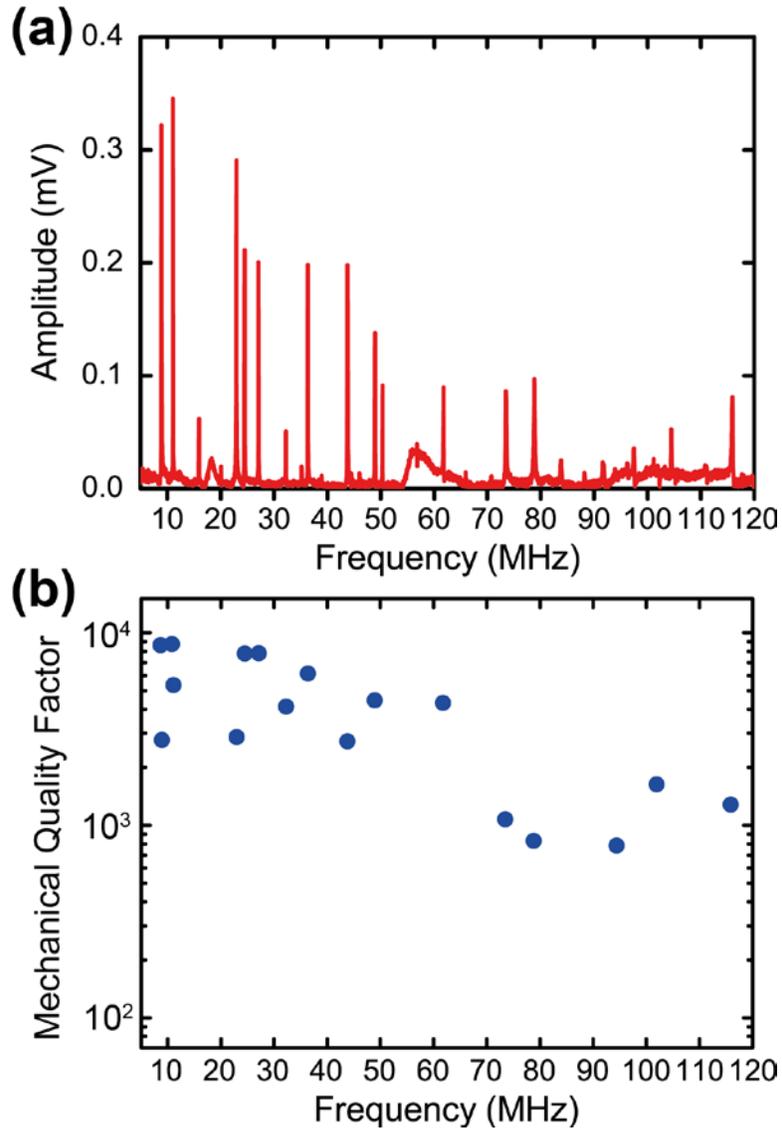

FIG. 2. Driven higher order mechanical resonances. (a) Acquired spectrum for one mechanical resonator. More than 20 mechanical eigenmodes in the range of 5 – 120 MHz can be excited via the electrodes and detected using the on-chip interferometer. (b) Extracted mechanical quality factors for resonances of two exemplary mechanical resonators of the same geometry (40 µm length and 600 nm beam width).

P. Rath, S. Ummethala, S. Diewald, G. Lewes-Malandrakis, D. Brink, N. Heidrich, C. Nebel, and W.H.P. Pernice

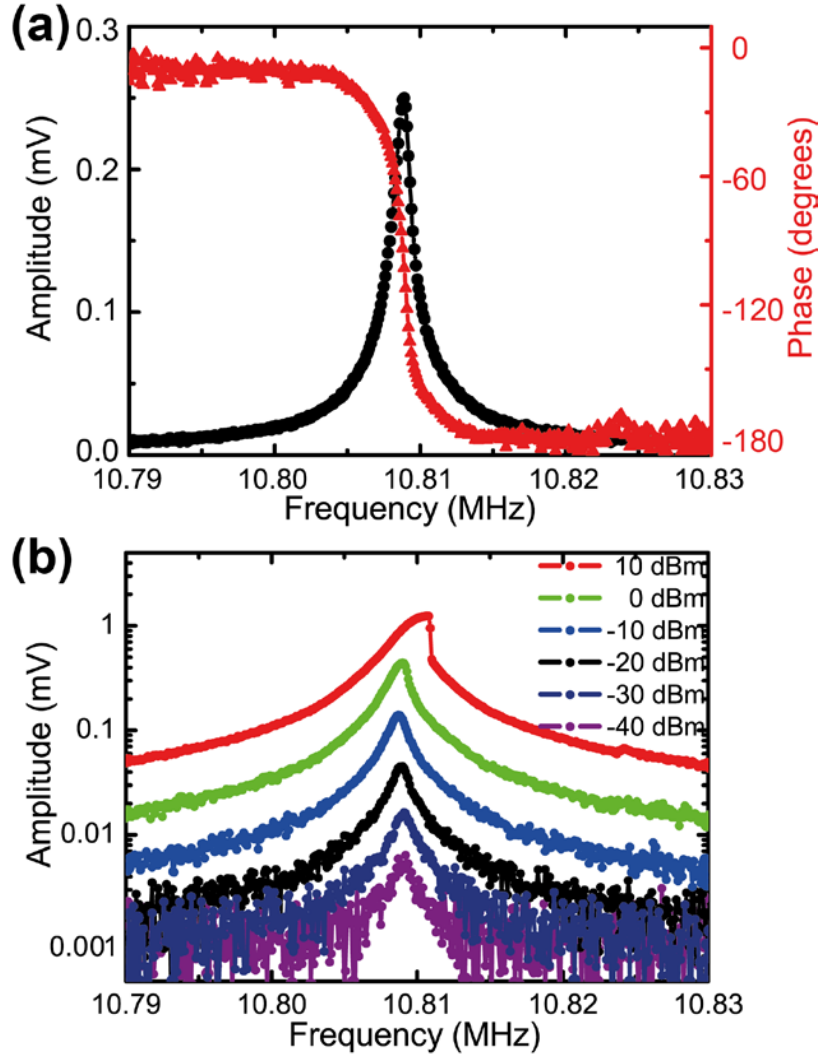

FIG. 3. Driven response for variable RF excitation power. (a) The amplitude (black) and phase (red) signal show the expected Lorentzian lineshape and a 180° phaseshift when sweeping the RF frequency across the resonance. (b) Driven mechanical resonances at a constant DC voltage of 5V and different RF driving powers. For RF powers above 0 dBm a stiffening Duffing non-linearity is observed, resulting from tensile stress in the diamond layer. The driven motion of the resonator can be detected for driving powers as low as -40 dBm.

P. Rath, S. Ummethala, S. Diewald, G. Lewes-Malandrakis, D. Brink, N. Heidrich, C. Nebel, and W.H.P. Pernice

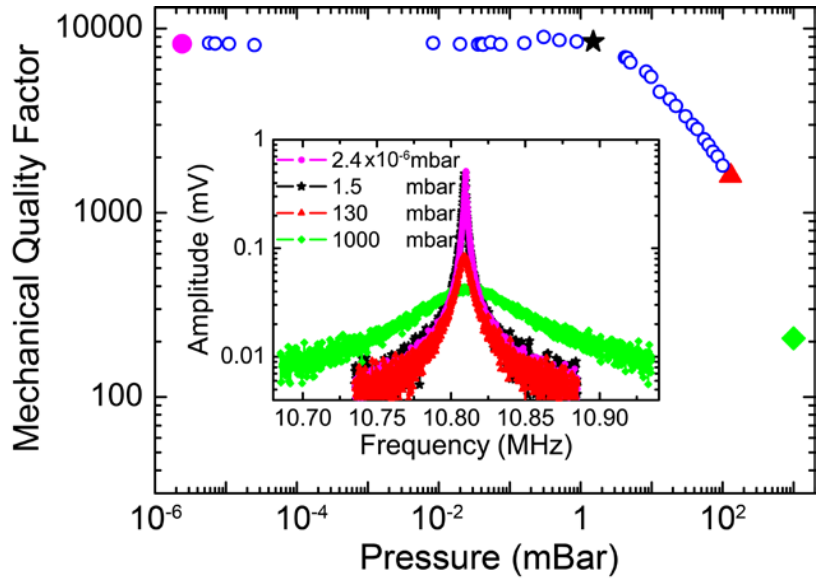

FIG. 4. Dependency of the quality factor on the pressure inside the vacuum chamber. Air damping is negligible for pressures from $10^{-6}$ mbar up to 1.5 mbar as the observed line shape and the amplitude of the mechanical resonance are the same (see corresponding spectra in the inset). The quality factor decreases from 8400 for pressures below 1.5 mbar to $Q = 210$ at ambient pressure.